\documentclass[conference]{IEEEtran}
\IEEEoverridecommandlockouts

\usepackage{amsmath,epsfig}
\usepackage{bm}
\usepackage{latexsym}
\usepackage{amssymb}
\usepackage{graphicx}

\def\bs{\mathbf{s}}
\def\bv{\mathbf{v}}
\def\bw{\mathbf{w}}
\def\bx{\mathbf{x}}
\def\by{\mathbf{y}}

\def\br{\mathbf{r}}
\def\bpsi{\bm{\psi}}

\begin{document}

\title{Cooperative Beamforming for
Wireless Ad Hoc Networks}

\author{\IEEEauthorblockN{Lun Dong, Athina P. Petropulu}
\IEEEauthorblockA{Department of Electrical and Computer Engineering\\
Drexel University, Philadelphia, PA 19104} \and \IEEEauthorblockN{H.
Vincent Poor}
\IEEEauthorblockA{School of Engineering and Applied Science\\
Princeton University, Princeton, NJ 08544}\thanks{This work was
supported by the National Science Foundation under Grants
ANI-03-38807, CNS-06-25637 and CNS-04-35052, and by the Office of
Naval Research under Grant N00014-07-1-0500.}}

\maketitle

\begin{abstract}
Via collaborative beamforming, nodes in a wireless network are
able to transmit a common message over long distances in an
energy efficient fashion.
 However, the process of making available the same message to all
 collaborating nodes introduces
delays. In this paper, a MAC-PHY cross-layer scheme is proposed that enables
collaborative beamforming at significantly reduced collaboration overhead. It
consists of two phases. In the first phase, nodes transmit locally in a random
access time-slotted fashion. Simultaneous transmissions from multiple source
nodes are viewed as linear mixtures of all transmitted packets. In the second
phase, a set of collaborating nodes, acting as a distributed antenna system,
beamform the received analog waveform to one or more faraway destinations. This
step requires multiplication of the received analog waveform by a complex
weight, which is independently computed by each cooperating node, and which
allows packets bound to the same destination to add coherently at the
destination node. Assuming that each node has access to location information,
the proposed scheme can achieve high throughput, which in certain cases exceeds
one. An analysis of the symbol error probability corresponding to the proposed
scheme is provided.

\end{abstract}

\section{Introduction}
Transmission over long distances often requires significant
amounts of energy in order to overcome attenuation. Energy is
usually a scarce commodity in wireless ad hoc networks, as
nodes typically operate on batteries, which in many cases are
difficult to replace or recharge. Thus, energy-efficient schemes for
long-distance transmission in wireless networks have recently
been of much interest. In some such situations, multihop may be
a preferred solution. However, there are several challenges in
transmitting real-time services over multiple hops. For
example, the traditional CSMA/CA based medium access control
(MAC) for avoiding collisions does not work well in a multihop
scenario because transmitters are often out of reach of other
nodes' sensing ranges. Thus, packets traveling across the
network experience interference and a large number of
collisions, which introduce long delays. Also, multihop
networks require a high node density which makes routing
difficult and affects the reliability of links \cite{Gharavi}.

Recently, a collaborative beamforming technique was proposed in
\cite{Ochiai}, in which randomly distributed nodes in a network
cluster form an antenna array and beamform data to a faraway
destination without each node exceeding its power constraint. The
destination receives data with high signal power. Beamforming with
antenna arrays is a well studied technology; it provides
space-division multiple access (SDMA) which enables significant
increases in communication rate. A challenge with implementing
beamforming in ad hoc networks is that the geometry of the network
may change dynamically. In \cite{Ochiai}, it was shown that randomly
distributed nodes can achieve a nice average beampattern with a
narrow main lobe and low side lobes. The directivity of the pattern
increases as the number of collaborating nodes increases. Such an
approach, when applied in the context of a multihop network reduces
the number of hops needed, thereby reducing packet delays and
improving throughput. However, to study network performance, one
must take into account the information-sharing time that is required
for node collaboration. If a time-division multiple-access (TDMA)
scheme were to be employed, the information-sharing time would
increase proportionally to the number of source nodes (i.e., the
nodes having packets to transmit).

In this paper we propose a scheme that is based on the idea of
collaborative beamforming, and reduces the time required for
information sharing. A preliminary version of the proposed scheme
appeared in \cite{CISS2007}. The work in this paper contains error
analysis that provides insight into the performance of the proposed
approach. The main idea is as follows. Different source nodes in the
network are allowed to transmit simultaneously. Collaborating nodes
receive linear mixtures of the transmitted packets. Subsequently,
each collaborating node transmits a weighted version of its received
signal. The weights are such that one or multiple beams are formed,
each focusing on one destination node, and reinforcing the signal
intended for a particular destination as compared to the other
signals. Each collaborating node computes its weight based on the
estimated channel coefficients between sources and itself. This
scheme achieves higher throughput and lower delay with the cost of
lower SINR as compared to \cite{Ochiai}. In the preliminary version
of this work \cite{CISS2007}, the analysis of interference at the
receiving node was done asymptotically, i.e., as the number of
collaborating nodes tends to infinity. Here we provide analytical
expressions for symbol error probability (SEP) that directly depend
on the number of collaborating nodes. The analysis shows how SEP is
affected by transmission power, signal-to-noise ratio, number of
simultaneously transmitting nodes and number of collaborating nodes.

\section{Background on collaborative beamforming} \label{back}

For simplicity, let us assume that sources and destinations are
coplanar. We index source nodes using a subscript $i,$ with
$t_i$ denoting the $i$-th node. At slot $n$, one source node
$t_m$ needs to transmit the signal $s_m(n)$ to a faraway
destination node $q_m$. Suppose that set of $N$ nodes,
designated as collaborating nodes $c_1,\ldots,c_N$, have access
to $s_m(n)$. The locations of these collaborating nodes follow
a uniform distribution over a disk of radius $R$. We denote the
location of $c_i$ in polar coordinates with respect to the
origin of the disk by $(r_i,\psi_i)$. Let $d_{im}(\phi_m)$, or simply
$d_{im}$, represent the distance between $c_i$ and the
destination $q_m$, where $\phi_m$ is the azimuthal angle of
$q_m$ with respect to the origin of the disk. $d_{0m}(\phi_m)$
or $d_{0m}$ denotes the distance between the origin of the disk
and $q_m$, so the polar coordinates of $q_m$ are
$(d_{0m},\phi_m)$. Moreover, let $d_{i}(\phi)$ denote the
distance between $c_i$ and some receiving point with polar
coordinate $(d_{0m}, \phi)$. The initial phases at the
collaborating nodes are set to
\begin{equation} \label{Psi1}
\Psi_i(\phi_m)=-\frac{2\pi}{\lambda} d_{im}(\phi_m), \ i=1,...,N
 \ .
\end{equation}
This requires knowledge of distances (relative to wavelength
$\lambda$) between nodes and destination, and applies to the
closed-loop case \cite{Ochiai}. Alternatively, the initial phase of
node $i$  can be
\begin{equation}
 \Psi_i(\phi_m)=\frac{2\pi}{\lambda} r_i \cos(\phi_m-\psi_i) \label{Psi2} \end{equation}
which requires knowledge of the node's position relative to some
common reference point, and corresponds to the open-loop case
 \cite{Ochiai}. In both cases synchronization is needed, which can
 be achieved via the use of the Global Positioning System (GPS).

The path losses between collaborating nodes and destination are
assumed to be identical for all nodes. The corresponding array
factor given the collaborating nodes at radial coordinates
${\br}=[r_1,...,r_N]$ and azimuthal coordinates
${\bpsi}=[\psi_1,...,\psi_N]$ at location with polar coordinate
$(d_{0m},\phi)$ is
\begin{eqnarray}
F(\phi;m|{\bf r,\bpsi})&=& \frac{1}{N}\sum_{i=1}^N
e^{j\Psi_i(\phi_m)} e^{j\frac{2\pi}{\lambda}d_i(\phi)} \ .
\end{eqnarray}

Under far-field assumptions, the array factor becomes \cite{Ochiai}
\begin{eqnarray} \label{100}
F(\phi;m|{\bf r,\bpsi})&=&\frac{1}{N}\sum_{i=1}^N  e^{j
\alpha(\phi;\phi_m)z_i} \label{2}
\end{eqnarray}
where  $\alpha(\phi;\phi_m)=4\pi (R/ \lambda)
\sin(\frac12(\phi_m-\phi))$,  and
$z_i=(r_i/R)\sin(\psi_i-\frac12(\phi_m+\phi))$. The random variable
$z_i$ has the following probability density function (pdf):
\begin{eqnarray}
f_{z_i}(z)=\frac{2}{\pi} \sqrt{1-z^2}, & \ -1\le z\le 1 \ .
\end{eqnarray}
Finally, the average beampattern can be expressed as \cite{Ochiai}
\begin{eqnarray} \label{101}
P_{\mathrm{av}}(\phi)&=& E_z\{ |F(\phi|{\bf z})|^2 \} \nonumber\\
&=& \frac{1}{N}+\left(1-\frac{1}{N}\right)\left|
2\frac{J_1(\alpha(\phi;\phi_m) )}{\alpha(\phi;\phi_m)} \right|^2
\end{eqnarray}
where $J_1(.)$ is the first-order Bessel function of the first kind.
When plotted as a function of $\phi$, $P_{\mathrm{av}}(\phi)$
exhibits a main lobe around $\phi_m$, and side lobes away from
$\phi_m$. It equals one in the target direction, and the sidelobe
level approaches $1/N$ as the angle moves away from the target
direction. The statistical properties of the beampattern were
analyzed in \cite{Ochiai}, where it was shown that under ideal
channel and system assumptions, directivity of order $N$ can be
achieved asymptotically with $N$ sparsely distributed nodes.

As we have noted, all of the collaborating nodes must have the same
information to implement beamforming. Thus, the source nodes need to
share their information symbols with all collaborating nodes in
advance. If a TDMA scheme were to be employed, the
information-sharing time would increase proportionally to the number
of source nodes. In the following, we propose a novel scheme to
reduce the information-sharing time and also allow nodes in the
network to transmit simultaneously.

\section{The proposed scheme} \label{proposed}
Here we refine the model of \cite{Ochiai}, focusing more directly on
the physical model for the signal, fading channel and noise. In
addition to the above assumptions, we will further assume the
following:

\begin{enumerate}
\item
 The network is divided into clusters, so that nodes in a cluster
can hear each other. In each cluster there is a node designated as
the cluster-head (CH). Nodes in a cluster do not need to transmit
their packets through the CH.

\item  A slotted packet system is considered, in which each packet
requires one slot for its transmission. Perfect synchronization is
assumed between nodes in the same cluster.

\item  Nodes transmit packets consisting of phase-shift keying (PSK) symbols each having the
same power $\sigma_s^2$. Also, nodes operate under half-duplex mode,
i.e., they cannot receive while they are transmitting.

\item Communication takes place over flat fading channels. The channel
gain during slot $n$ between source $t_i$ and collaborating node
$c_j$ is denoted by $a_{ij}(n)$. It does not change within one slot,
but can change between slots.  The channel gains are
 independent and identically distributed (i.i.d.) complex Gaussian random variables with
zero means and variances $\sigma_a^2$ across both time and space,
i.e., $a_{ij}(n) \sim \mathcal{CN}(0, \sigma_a^2)$.

\item  The complex baseband-equivalent channel gain between
    nodes $c_i$ and $q_m$ is $ b_{im} e^{{j\frac{2\pi
    }{\lambda}}d_{im}} $ \cite{tse-book}, where $b_{im}$ is
    the path loss. The
distances between collaborating
    nodes and destinations are much greater than the
    maximum distance between source and collaborating
    nodes. Thus, $b_{im}$ is assumed to be identical for
    all collaborating nodes and equals the path loss
    between the origin of the disk and the destination,
    denoted by $b_m$.

\end{enumerate}

Suppose that cluster $C$ contains $J$ nodes. During slot $n$, source
nodes $t_1,\ldots, t_K$ need to communicate with nodes
$q_1,\ldots,q_K$ that belong to clusters $C_1,\ldots,C_K$,
respectively. The azimuthal angle of destination $q_i$ is denoted by
$\phi_i$. The packet transmitted by node $t_j$ consists of $L$
symbols $\bs_j(n) \triangleq [s_j(n;0), \ldots, s_j(n;L-1)]$. Due to
the broadcast nature of the wireless channel, non-source nodes in
cluster $C$ hear a collision, i.e., a linear combination of the
transmitted symbols. More specifically, node $c_i$ hears the signal
\begin{equation} \label{crec}
\bx_i(n)=\sum_{j=1}^K a_{ji}(n)\bs_j(n) + \bw_i(n)
\end{equation}
where $\bw_i(n)=[w_i(n;0),\ldots, w_i(n;L-1)]$ represents noise
 at the receiving node $c_i$.
 The noise is assumed to be of zero mean and with covariance matrix
 $\sigma_w^2{\bf I}_L$, where ${\bf I}_L$ is an $L \times L$ identity matrix.

Once the CH establishes that there has been a transmission, it
initiates a collaborative transmission period (CTP), by sending a
control bit to all nodes, e.g., $1$, via an error-free control
channel. The CH will continue sending a $1$ in the beginning of each
subsequent slot until the CTP has been completed. The cluster nodes
cannot transmit new packets until the CTP is over.

Let $q_m$ denote the destination of $\bs_m(n)$. In slot $n+m, \
m=1,\ldots ,K$, each collaborating node $c_i$ transmits the signal
\begin{equation}
 \tilde \bx_i(n+m)=\bx_i(n) \mu_m a^*_{mi}(n) e^{\Psi_i(\phi_m)}
\end{equation}
where $\mu_m$ is a scalar used to adjust the transmit power and is
the same for all collaborating nodes.  $\mu_m$ is of the order of
$1/N$.

Collaborating nodes need to know which are source nodes and then
estimate the channel between all source nodes and themselves. One
possible way to implement this is to use orthogonal IDs, as
discussed in \cite{CISS2007}, \cite{alliances}.

Also, collaborating nodes require the knowledge of their initial
phases. In closed-loop mode, each collaborating node can
independently synchronize itself to a beacon sent from the
destination and adjusts its initial phase to it \cite{Ochiai}. In
open-loop mode, each collaborating node needs to know its relative
position from a predetermined reference point (e.g. the origin of
the disk) within the cluster, which can be achieved by the use of
GPS. To obtain initial phases, collaborating nodes also require
knowledge of the azimuths of the destinations so that the beams can
be steered toward desired directions, which may be broadcast by the
CH via a control channel.

Given the collaborating nodes at radial coordinates
${\br}=[r_1,...,r_N]$ and azimuthal coordinates
 ${\bpsi}=[\psi_1,...,\psi_N]$, the
received signal at an arbitrary location with polar coordinates
$(d_{0m},\phi)$, is
\begin{equation} \label{rec}
\by(\phi;m|{\br,\bpsi})= \sum_{i=1}^N b_m \tilde \bx_i(n+m) e^{j
\frac{2\pi}{\lambda}d_i(\phi)} + \bv(n+m)
\end{equation}
where $\bv(n+m)$ represents noise at the receiver during slot $n+m$.
The covariance matrix of $\bv(n+m)$ equals
  $\sigma_v^2{\bf I}_L$.

It was shown in \cite{CISS2007} that, as $N\rightarrow \infty$ and
omitting the noise, $\by(\phi_m;m|{\bf r,\bpsi}) \rightarrow N \mu_m
b_m \sigma_a^2 \bs_m(n)$. Thus, the destination node $q_m$ receives
a scaled version of $\bs_m(n)$. The beamforming step is completed in
$K$ slots, reinforcing one source signal at a time.

Assuming that all of the $K$ source packets have distinct
destinations at different resolvable directions, multiple beams can
be formed in one slot, each beam focusing on one direction and
reinforcing one source signal. In the rest of the paper, for
simplicity we will consider only the case in which a single beam is
formed during slot $n+m$, focusing on destination $q_m$. The results
obtained under this assumption can be readily extended to multiple
simultaneous beams.

Taking into account the assumptions on channels and noise,
 the average beampattern was derived in \cite{CISS2007}.
 Defining the throughput, $T$, as the average number of
packets that are successfully transmitted in a time slot, we showed
in \cite{CISS2007} that $K/(1+K) \le T\le K/2$, which could be
greater than 1. Also, in \cite{CISS2007}, we showed that under a
fixed transmit power, the average signal-to-interference plus noise
ratio (SINR) is asymptotically $\beta'$ times less than that of
\cite{Ochiai}, where
$\beta'=K+1+\frac{\sigma_w^2}{\sigma_s^2\sigma_a^2}$.

\section{Symbol Error Probability (SEP)}
\label{ser}

In the following, for simplicity  we omit the time index, and
replace $\by$, $\tilde{\bx}_i$, $\bx_i$, $\bs_i$, $\bw_i$ and $\bv$
in the above equations by $y$, $\tilde{x}_i$, $x_i$, $s_i$, $w_i$
and $v$ (i.e., with one of their samples) respectively.

Our analysis will be conditioned on $K$, the number of
simultaneously transmitting nodes.  In general, $K$ is a random
variable, whose distribution is a function of the traffic
characteristics, e.g, traffic load, traffic distribution,
transmission control scheme, etc. In the simple case in which each
node transmits with identical probability $P_t$, $K$ has a binomial
distribution. Once the distribution of $K$ is given then we can
determine the SEP as $P_s=\sum_{K=1}^{J}P(K)P_s(K) $.

From (\ref{rec}), the received signal at the destination $q_m$ is
\begin{eqnarray}
&& y(\phi_m;m)= \mu_m b_m \sum_{i=1}^N |a_{mi}|^2 s_m \nonumber \\
& &  \ \ \ \ + \mu_m b_m \sum_{i=1}^N a_{mi}^* (\sum_{j=1 \atop
j\neq m}^{K}{a_{ji} s_{j}}+w_{i}) +v
\end{eqnarray}
where the first term is the desired signal and the remaining terms
represent interference and noise. Recall that $a_{ji} \sim
\mathcal{CN}\left(0, \sigma_a^2\right)$. Since $s_j$ is a PSK
symbol, the magnitude of $a_{ji}s_j$ is $\sigma_s^2 |a_{ji}|$ and
its phase is still uniformly distributed in $[0, 2\pi]$. Thus,
$a_{ji}s_j \sim \mathcal{CN}\left(0, \sigma_a^2\sigma_s^2 \right)$.
Therefore,
\begin{equation}
\eta_i \triangleq \sum_{j=1 \atop j\neq m}^{K}{a_{ji} s_{j}}+w_{i}
\sim \mathcal{CN}\left(0, \sigma_\eta^2\right)
\end{equation}
where $\sigma_\eta^2\triangleq(K-1)\sigma_a^2\sigma_s^2+\sigma_w^2$.

Given $a_{mi}$, the instantaneous SINR, $\gamma$, equals
\begin{equation} \label{SNR}
\gamma= \frac{\mu_m^2 b_m^2 (\sum_{i=1}^N |a_{mi}|^2)^2
\sigma_s^2}{\mu_m^2 b_m^2 \sum_{i=1}^N |a_{mi}|^2
\sigma_\eta^2+\sigma_v^2}= \frac{\mu_m^2 b_m^2 \xi^2
\sigma_s^2}{\mu_m^2 b_m^2 \xi \sigma_\eta^2+\sigma_v^2}
\end{equation}
where $\xi \buildrel \triangle \over = \sum_{i=1}^N |a_{mi}|^2$.

Note that $\mu_m$ is of  order  $1/N$. As $N \rightarrow \infty$,
$\mu_m^2 b_m^2 \xi \sigma_\eta^2 \rightarrow 0$, and $\gamma$
reduces to $\mu_m^2 b_m^2 \xi^2 \sigma_s^2/\sigma_v^2$, which
corresponds to the scenario of additive white Gaussian noise
(AWGN). Thus, under certain transmit powers, no matter how large $N$
is, the SEP of the proposed scheme is always lower bounded by the
SEP under AWGN.

Since $|a_{mi}|$ is Rayleigh distributed, $\xi \sim \mathrm{Erlang}
(N, \sigma_a^2) $. The pdf of the Erlang distribution is
\begin{eqnarray} \label{erlang}
\mathrm{Erlang} (k, \theta):& f(x;k,\theta)=\frac{x^{k-1}
e^{-\frac{x}{\theta}}}{\theta^k(k-1)!}, & x\geq 0 \ .
\end{eqnarray}

The moment generating function (MGF) of $\gamma$ is
\begin{eqnarray}
\mathcal{M}_\gamma(s)&=& \int_{-\infty}^{\infty} \exp (s \gamma)
f_{\xi}(\xi) d \xi \nonumber \\
&=&  \int_0^\infty \exp (\frac{s \mu_m^2 b_m^2 \xi^2
\sigma_s^2}{\mu_m^2 b_m^2 \xi \sigma_\eta^2+\sigma_v^2})
\frac{\xi^{N-1}e^{-\frac{\xi}{\sigma_a^2}}}{\sigma_a^{2N} (N-1)!} d
\xi \nonumber \\& &
\end{eqnarray}
based on which, the average SEP for M-PSK symbols is
\cite{simon-book}
\begin{eqnarray} \label{exactSEP}
&&P_s(K)=\frac{1}{\pi}\int_{0}^{\frac{(M-1)\pi}{M}}
\mathcal{M}_\gamma\left(-\frac{\sin^2(\pi/M)}{\sin^2\varphi}\right)
d \varphi \nonumber\\
&=& \frac{1}{\pi}\int_{0}^{\frac{(M-1)\pi}{M}} \int_0^\infty \exp
\left(-\frac{\sin^2(\pi/M)}{\sin^2\varphi} \cdot \frac{ \mu_m^2
b_m^2 \xi^2 \sigma_s^2}{\mu_m^2 b_m^2 \xi
\sigma_\eta^2+\sigma_v^2}\right)
\nonumber\\
&& \times \frac{\xi^{N-1}e^{-\frac{\xi}{\sigma_a^2}}}{\sigma_a^{2N}
(N-1)!} d \xi d \varphi \ .
\end{eqnarray}

Since there is no closed-form  expression for
$\mathcal{M}_\gamma(s)$ or $P_s(K)$, in the following we will make
some approximations to simplify the above expressions.

\subsection{A Simple Bound for SEP} \label{simple_bounds}

Let us fix an $\epsilon > 0$, and define $\xi_0$ such that $P(\xi
\leq \xi_0)=\epsilon$. Also, let us define
\begin{equation}
\tilde{\gamma} \triangleq \frac{\mu_m^2 b_m^2 \xi^2
\sigma_s^2}{\mu_m^2 b_m^2 \xi \sigma_\eta^2+\sigma_v^2 \xi/\xi_0}
=\frac{\mu_m^2 b_m^2 \sigma_s^2}{\mu_m^2 b_m^2
\sigma_\eta^2+\sigma_v^2/\xi_0} \cdot \xi  \triangleq
c_{\tilde{\gamma}} \xi \ .
\end{equation}
When $\epsilon$ is small, it holds with probability $\ge 1-\epsilon$
that  $\tilde{\gamma} \leq \gamma$. Since $c_{\tilde{\gamma}}>0$ and
$s$ is negative in the range of interest, we can always find a small
enough $\epsilon$ so that $\mathcal{M}_{\tilde{\gamma}} (s) \geq
\mathcal{M}_\gamma(s)$.

Note that $\tilde{\gamma} \sim \mathrm{Erlang}(N,\sigma_a^2
c_{\tilde{\gamma}} )$ and thus the MGF of $\tilde{\gamma}$ is of the
following simple form:
\begin{equation}
\mathcal{M}_{\tilde{\gamma}}(s)=(1-s \sigma_a^2
c_{\tilde{\gamma}})^{-N} \ .
\end{equation}

From  (\ref{exactSEP}), the SEP for M-PSK symbols based on
$\tilde{\gamma}$ is
\begin{equation}  \tilde{P}_s(K) =
\frac{1}{\pi}\int_{0}^{(M-1)\pi/M}
(1+\frac{\sin^2(\frac{\pi}{M})\sigma_a^2
c_{\tilde{\gamma}}}{\sin^2\varphi})^{-N} d \varphi \ .
\end{equation}

Defining $c\triangleq \sin^2(\frac{\pi}{M})\sigma_a^2
c_{\tilde{\gamma}}$, and using the result of Eq. (5A. 17) in
\cite{simon-book} we obtain
 {
\setlength\arraycolsep{0.1em}
\begin{eqnarray} \label{approxSEP} && \tilde{P}_s(K) =
\frac{1}{\pi}\int_{0}^{(M-1)\pi/M} (1+\frac{c}{\sin^2\varphi
})^{-N} d \varphi \nonumber\\
&=&
\frac{M-1}{M}-\frac{1}{\pi}\sqrt{\frac{c}{1+c}}\{(\frac{\pi}{2}+\tan^{-1}\zeta
)\sum_{n=0}^{N-1}\left(2n \atop n \right)
\frac{1}{[4(1+c)]^n}\nonumber\\
&&+\sin(\tan^{-1} \zeta)
\sum_{n=1}^{N-1}\sum_{j=1}^{n}\frac{T_{jn}}{(1+c)^n}[\cos(\tan^{-1}\zeta)]^{2(n-j)+1}\}
\nonumber \\
\end{eqnarray}}
where
\begin{eqnarray}
\zeta \triangleq \sqrt{\frac{c}{1+c}} \cot
\left(\frac{\pi}{M}\right)
\end{eqnarray}
and
\begin{eqnarray}
T_{jn} \triangleq \frac{\left(2n \atop n \right)}{\left(2(n-j) \atop
n-j \right)4^j[2(n-j)+1]} \ .
\end{eqnarray}

Recalling that $\mathcal{M}_{\tilde{\gamma}} (s)  \geq
\mathcal{M}_\gamma(s) \geq 0$, we have $\tilde{P}_s(K)\geq P_s(K)$.
The result of (\ref{approxSEP}) is an upper bound of the exact SEP
of (\ref{exactSEP}).

An even simpler upper bound for $\tilde{P}_s(K)$ can be obtained
based on Eq. (5A.76) of \cite{simon-book}:
\begin{eqnarray}
\label{upper} \tilde{P}_s(K) &\leq & \frac{M-1}{M}
\left(1+\frac{c}{\sin^2(\frac{(M-1)\pi}{M})}\right)^{-N}
\end{eqnarray}
which for BPSK becomes
\begin{eqnarray} \label{BPSKupper}
\tilde{P}_s(K) &\leq & \frac{1}{2} \left(1+\sigma_a^2
c_{\tilde{\gamma}}\right)^{-N} \ .
\end{eqnarray}

\textbf{Remark:} As $\mu_m \rightarrow \infty$, $c_{\tilde{\gamma}}$
reduces to $\sigma_s^2/\sigma_\eta^2$, in which case the
corresponding result of (\ref{approxSEP}) can be viewed as a lower
bound when the transmit power of collaborating nodes approaches
infinity. It shows that no matter how large the transmit power is,
the SEP can never be smaller than this bound. The SEP floor is a
result of the interference from other source nodes. To achieve lower
SEP for a given $K$, one must increase $N$. Based on (\ref{upper}),
this bound decreases approximately in a power-law fashion as  $N$
increases.

\section{Simulations}
\label{sim}

In this section, we study the SEP performance of the proposed method
via simulations, and also via the proposed analytical expressions.

We assume the channels among nodes in a cluster are selected from
zero-mean complex Gaussian processes, which are constant within one
slot, but vary between slots. Let us define $\gamma_1 \triangleq
\sigma_s^2\sigma_a^2/\sigma_w^2,$ which represents the average SNR
in the process of information sharing, and define $\gamma_2
\triangleq N^2 \mu_m^2 b_m^2 \sigma_s^2 \sigma_a^4/ \sigma_v^2$ to
represent the asymptotic average SNR (when $N \rightarrow
\infty$) at the receiver. Note that $\gamma_2$ is  independent of
$N$ since $\mu_m$ is of the order of $1/N$. Eq. (\ref{SNR}) can be
rewritten by
\begin{eqnarray}
\gamma=\frac{\tilde{\xi}^2/N^2}{\frac{K-1+\gamma_1^{-1}}{N^2}
\tilde{\xi}+\gamma_2^{-1}}
\end{eqnarray}
where $\tilde{\xi}=\xi/\sigma_a^2 \sim \mathrm{Erlang}(N,1)$. Then,
the SINR is determined only by $\gamma_1$, $\gamma_2$, $K$ and $N$.
Each packet contains BPSK symbols, so SEP is equivalent to BER. We
take $\epsilon= 0.01$. Also, we assume perfect knowledge of
channels, number of source nodes and destination information. Only
one beampattern is formed in each slot. For simulation-based BER, we
perform a Monte-Carlo experiment consisting of $10^6$ repeated
independent trials.

\begin{figure}[htb]
 \centerline{\epsfig{figure= 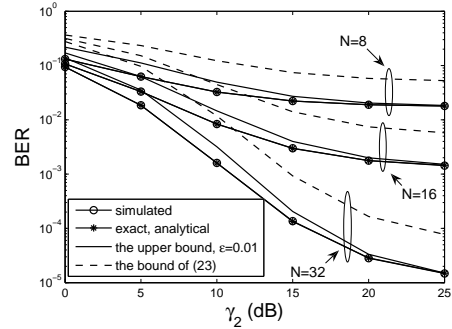,width=6.5cm}}
\caption{BER vs. $\gamma_2$ ($K=4,\gamma_1=20$ dB); $N = 8,16,32$;
empirical results, analytical exact results and upper bounds.}
 \label{BER_gamma2}
\end{figure}

Fig. \ref{BER_gamma2} shows the BER versus $\gamma_2$ estimated from
the network simulation ($\circ$ line) when $ K=4$ nodes transmit
all the time. The parameter $\gamma_1$ is fixed at 20 dB. The
estimated BER is in perfect agreement with the analytical result for
the exact SEP of (\ref{exactSEP})(``$\ast$'' line); in fact the two
lines are indistinguishable. The upper bound on the exact SEP,
computed by (\ref{approxSEP}), is shown as the solid line. One can
see that $\epsilon=0.01$ can  guarantee a tight bound under various
parameters and SNR ranges. The simple upper bound computed via
(\ref{BPSKupper}) is also shown (dashed lines).

Extensive simulations confirm that the simulation-based BER and
analytical SEP  match well under a wide variety of scenarios. Thus,
in the following we will simply use the analytical result of
(\ref{exactSEP}) to study the performance of the proposed method.

\begin{figure}[htb]
 \centerline{\epsfig{figure= 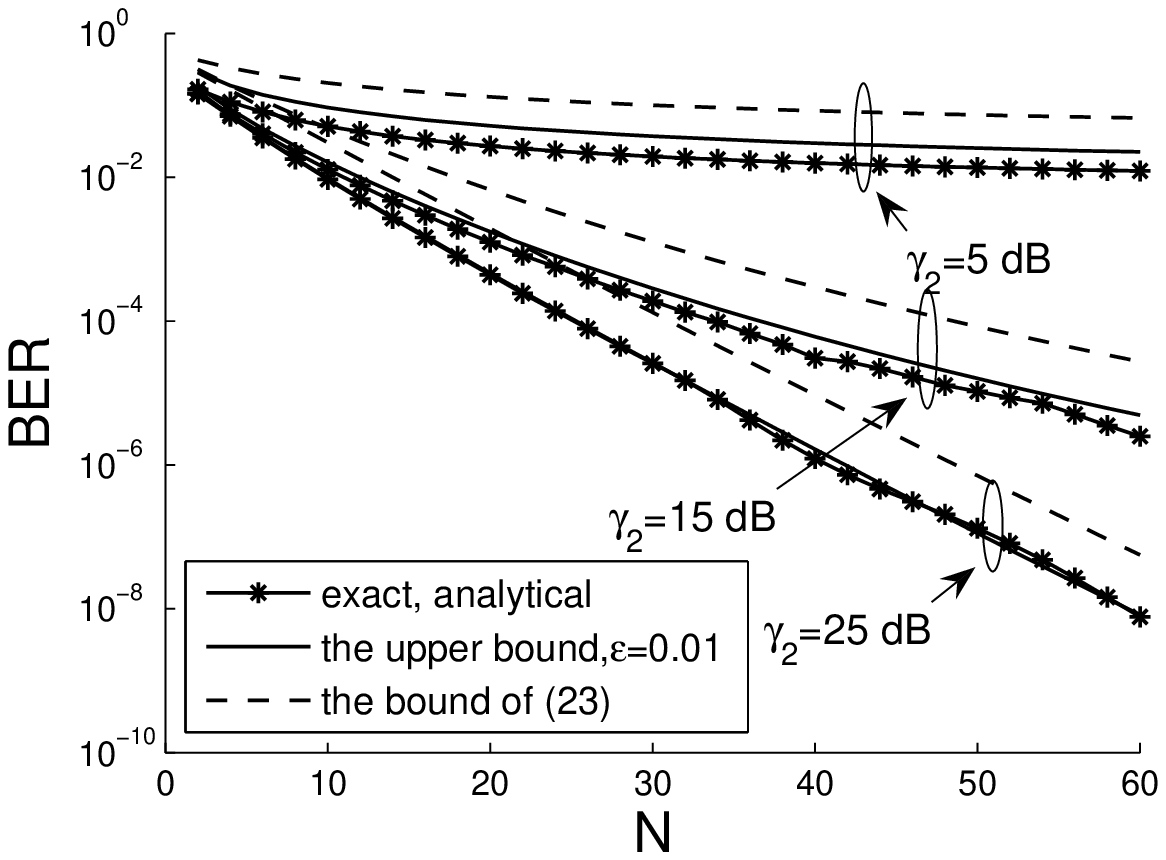,width=6.5cm}}
\caption{BER vs. $N$ ($K=4,\gamma_1=20$ dB); $N = 8,16,32$;
analytical exact results and upper bounds.}
 \label{BER_N}
\end{figure}

\begin{figure}
 \centerline{\epsfig{figure= 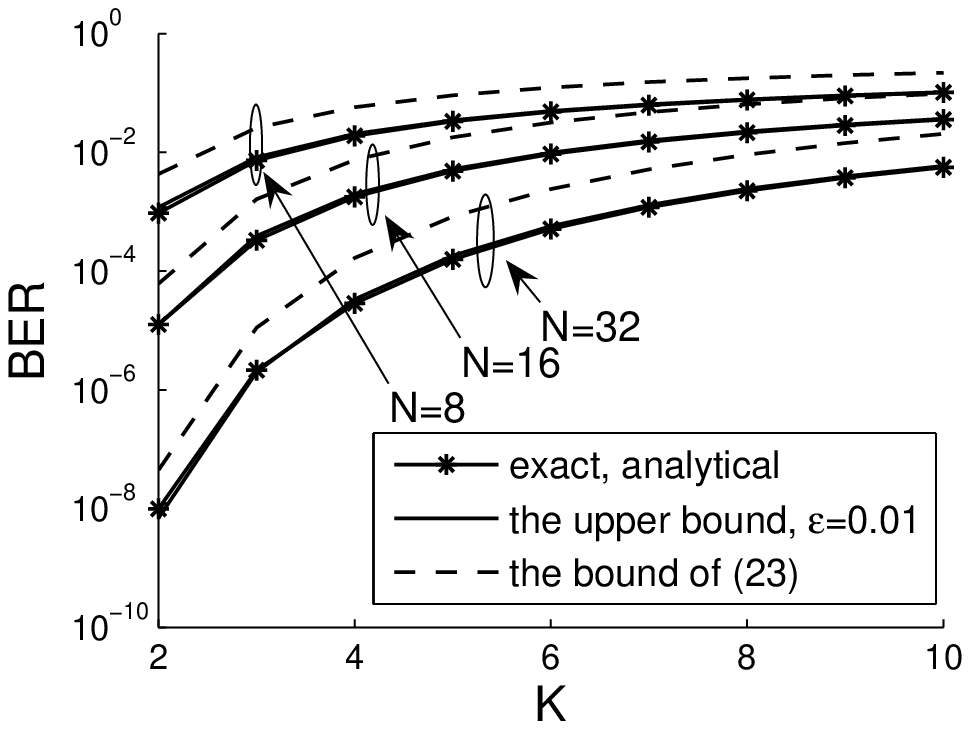,width=6.5cm}}
\caption{BER vs. $K$ ($\gamma_1=\gamma_2=20$ dB); $N = 8,16,32$;
analytical exact results and upper bounds.}
 \label{BER_K}
\end{figure}

\begin{figure}
 \centerline{\epsfig{figure= 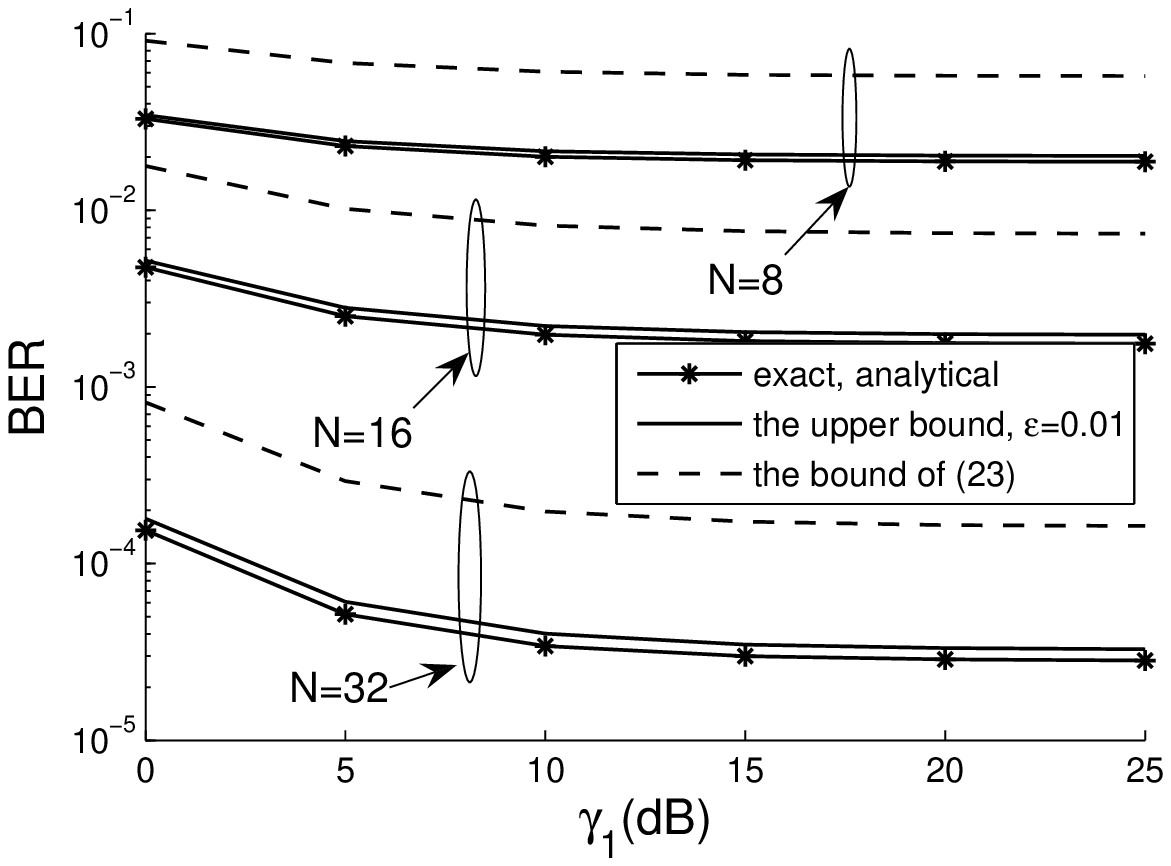,width=6.5cm}}
\caption{BER vs. $\gamma_1$ ($K=4, \gamma_2=20 $dB); $N = 8,16,32$;
analytical exact results and upper bounds.}
 \label{BER_gamma1}
\end{figure}

Fig. \ref{BER_N} shows how the BER depends on the number of
collaborating nodes for  $\gamma_1= 20$ dB and different values of
$\gamma_2$. Fig. \ref{BER_K} shows how $K$ affects BER, where
$\gamma_1=\gamma_2= 20$ dB. As $K$ increases, the SEP increases.
Fig. \ref{BER_gamma1} shows how BER changes with $\gamma_1$, where
$\gamma_2= 20$ dB and $K=4$. Recall that
$\sigma_\eta^2=\sigma_a^2\sigma_s^2(K-1+\gamma_1^{-1})$. $K$ plays a
dominant role in the interference (when $K>1$). As observed in Fig.
\ref{BER_gamma1}, the SEP decreases only slightly with the increase
of $\gamma_1$.

\section{Conclusions} \label{con}

We have proposed a scheme for wireless ad hoc networks that uses the
idea of collaborative beamforming and at the same time
 reduces the time needed for information sharing during the
collaborative phase. We have provided an analysis of the SEP, which
shows how the performance depends on the number of collaborating
nodes, the number of simultaneously source users and noise levels at
collaborating nodes and the final destination node.

\footnotesize{
}

\end{document}